\documentclass[aps,prl,floatfix,showpacs,twocolumn,tightenlines]{revtex4}

\usepackage{color}
\usepackage{graphicx}
\usepackage{latexsym}

\begin{document}

\title{Measuring Entangled Qutrits and Their Use for Quantum Bit Commitment}

\author{N.~K.~Langford}
\email{langford@physics.uq.edu.au}
\affiliation{Department of Physics, University of Queensland, Brisbane, QLD 4072, Australia}

\author{R.~B.~Dalton}
\affiliation{Department of Physics, University of Queensland, Brisbane, QLD 4072, Australia}

\author{M.~D.~Harvey}
\affiliation{Department of Physics, University of Queensland, Brisbane, QLD 4072, Australia}

\author{J.~L.~O'Brien}
\affiliation{Department of Physics, University of Queensland, Brisbane, QLD 4072, Australia}

\author{G.~J.~Pryde}
\affiliation{Department of Physics, University of Queensland, Brisbane, QLD 4072, Australia}

\author{A.~Gilchrist}
\affiliation{Department of Physics, University of Queensland, Brisbane, QLD 4072, Australia}

\author{S.~D.~Bartlett}
\affiliation{Department of Physics, University of Queensland, Brisbane, QLD 4072, Australia}

\author{A.~G.~White}
\affiliation{Department of Physics, University of Queensland, Brisbane, QLD 4072, Australia}

\date{\today}

\begin{abstract}
We produce and holographically measure entangled qudits encoded in transverse spatial modes of single photons. With the novel use of a quantum state tomography method that only requires two-state superpositions, we achieve the most complete characterisation of entangled qutrits to date. Ideally, entangled qutrits provide better security than qubits in quantum bit-commitment: we model the sensitivity of this to mixture and show experimentally and theoretically that qutrits with even a small amount of decoherence cannot offer increased security over qubits.
\end{abstract}


\maketitle

Many two-level quantum systems, or \emph{qubits}, have been used to encode information \cite{QIC01}; using $d$-level systems, or \emph{qudits}, enables access to larger Hilbert spaces, which can provide significant improvements over qubits such as increased channel capacity in quantum communication \cite{FujiwaM1}. When entangled, \emph{qutrits} ($d$=3) provide the best known levels of security in quantum bit-commitment and coin-flipping protocols, which cannot be matched using qubit-based systems \cite{SpekkeRW1}. The ability to completely characterise entangled qudits is critical for applications. This is only possible using quantum state tomography \cite{JamesDFV1,ThewRT1}.

\begin{figure}[b!]
\includegraphics[width=80mm]{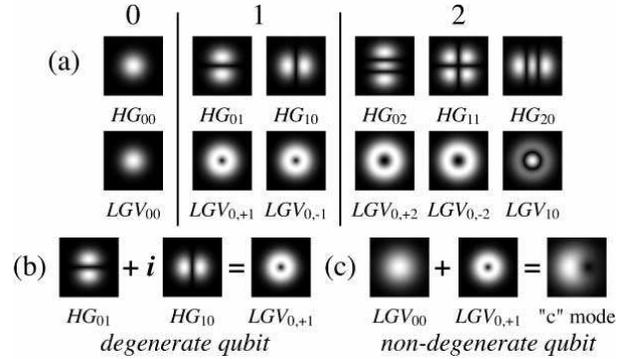}
\caption{(a) The first three orders of two paraxial mode families: the Hermite-Gauss modes ($HG_{rs}$), with $r$ horizontal and $s$ vertical lines of phase discontinuity; the Laguerre-Gauss-Vortex modes ($LGV_{pl}$), with $p$ ring phase discontinuities and a charge $l$ phase singularity, or vortex. The mode order is $r$+$s$ for $HG_{rs}$ modes and $2p$+$l$ for $LGV_{pl}$ modes. Superposition states for (b) degenerate and (c) non-degenerate qubits, where the logical modes are respectively of the same and different orders. The displaced singularity in the non-degenerate qubit moves around the beam centre as it propagates.}
\label{spatialQI}
\end{figure}

Entangled qudits have been realised in few physical systems, and only indirect measurements have been made of the quantum states of these systems. Qutrit entanglement has been generated between the arrival times of correlated photon pairs, where fringe measurements were used to infer features such as fidelities with specific entangled states and to estimate a potential Bell violation \cite{ThewRT2}. It is also possible to encode qudits in the transverse spatial modes of a photon, Fig.~\ref{spatialQI}. There have been measurements demonstrating, but again not quantifying, spatial mode entanglement in parametric downconversion \cite{MolinaG1}, including fringe measurements \cite{MairA1,VaziriA3} and the violation of a two-qutrit Bell inequality \cite{VaziriA1,note0}.

Here, we use quantum state tomography to completely characterise entangled, photonic qudits (both $d=2$ and $3$) encoded in transverse spatial modes, measuring the amount of entanglement and the degree of mixture. We show how to use the qutrit system in a quantum bit-commitment protocol and investigate the experimental requirements for achieving the best known security \cite{SpekkeRW1}. To illustrate these results, we first introduce and demonstrate two conceptually distinct ways of encoding information in transverse spatial modes, which differ in the behaviour of  superposition states. This work constitutes the most complete characterisation of spatially-encoded qubits and qutrits and the first quantitative measurement of entangled qutrit states.

The Gaussian spatial modes are a complete basis for describing the paraxial propagation of light \cite{Siegman}. Two orthonormal mode families are shown in Fig.~\ref{spatialQI}(a): the Hermite-Gauss ($HG_{rs}$) and Laguerre-Gauss-Vortex ($LGV_{pl}$). These modes are self-similar under propagation; modes of the same {\em order} experience the same propagation-dependent phase shift, the Gouy phase shift. We define \emph{degenerate} qudits to be constructed from basis states of the same order [Fig.~\ref{spatialQI}(b)]. Conversely, \emph{non-degenerate} qudits contain states of different orders [Fig.~\ref{spatialQI}(c)]; the different Gouy phases cause non-degenerate qudit superpositions to change phase as they propagate.

\begin{figure}[b!]
\includegraphics[width=80mm]{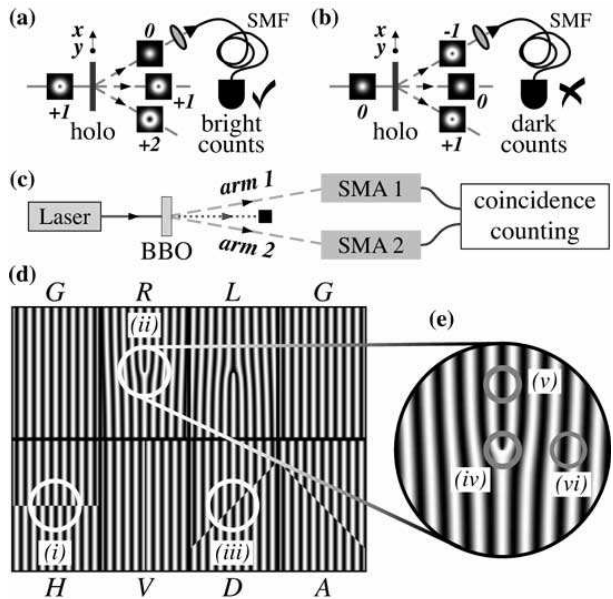}
\caption{Quantum state tomography of spatial modes. A spatial mode analyser (SMA) realised with a $LGV_{0,+1}$ hologram: (a) the target mode ($LGV_{0,+1}$) couples into a single-mode fibre; (b) other modes (e.g.\ $LGV_{00}$) are rejected. The images are labelled with the charge of the phase singularity in the beam. (c) Conceptual layout for tomography. Two SMAs analyse the mode of the energy degenerate pairs (820$\pm$10 nm), postselected by counting in coincidence for 100 s with fibre-coupled avalanche photodiodes ($\sim 100$ counts/s). (d) The 8-segment analysis hologram \cite{holoeff} used in all our experiments: the labels ($G$, $R$, {\em etc.}) correspond to the main spatial mode analysed by that segment. The positions (\emph{i}-\emph{iii}) for (d) degenerate and (\emph{iv}-\emph{vi}) for (e) non-degenerate qubits correspond, respectively, to measuring one computational basis state and the two equal superposition states, $(|0\rangle+{\rm i}|1\rangle)/\sqrt{2}$ and $(|0\rangle+|1\rangle)/\sqrt{2}$.}
\label{dcQST}
\end{figure}

When encoding in photon polarisation, the quantum state is manipulated with wave plates and selected using a polarising beam splitter \cite{WhiteAG2}. In spatial encoding, the wave-plate function is achieved with a hologram, and the beam-splitter with a single-mode fibre (SMF), which selects the lowest order spatial component ($HG_{00} \!\equiv\! LGV_{00}\!\equiv\! G$) and interferometrically rejects all higher order modes. A spatial mode analyser (SMA) combines these two components with a detector. The hologram first converts the target mode into the mode $G$, which is then selected by the fibre [Fig.~\ref{dcQST}(a)]. All other modes are rejected [Fig.~\ref{dcQST}(b)] with typical extinctions of $\sim$$10^{-3}$ --- equivalent to standard commercial polarising beam splitters. We use different holograms to measure different states, as described below.

Quantum state tomography requires a series of complementary measurements on a large ensemble of identically prepared copies of the system \cite{JamesDFV1}. Rather than measure $d$-state superpositions \cite{ThewRT1}, we choose a set of measurements constructed from only basis states, $|j \rangle$,  and two-state superpositions, $|p^{+} \rangle$ and $|q^{+} \rangle$, where $|p^{\pm} \rangle$=$(|j \rangle$$\pm$$|k \rangle)/\sqrt{2}$, $|q^{\pm} \rangle$=$(|j \rangle$$\pm$$i |k \rangle)/\sqrt{2}$, and $j,k \in \{0,1,...d-1\}$ \cite{map}. In practice, we use an over-complete set including $|p^{-} \rangle$ and $|q^{-} \rangle$, which allows more accurate normalisation when converting the data to measurement probabilities. We obtain a physical density matrix using an optimisation procedure \cite{JamesDFV1}; the over-specification also makes the optimisation less sensitive to outlying data points. Using this two-state tomographic technique, we characterise the output from a Type-I down-conversion source pumped by a blue diode laser [Fig.~\ref{dcQST}(c)]. The two SMAs image partial, banana-shaped sections of the cone of energy degenerate photon pairs and so see significant contributions from spatial components other than $G$.

\begin{figure}[b!]
\includegraphics[width=80mm]{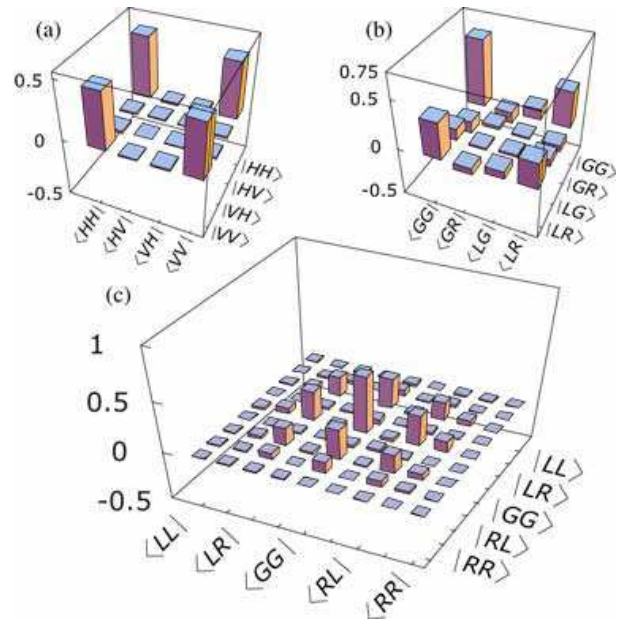}
\caption{Measured density matrices (real parts) for: (a) entangled degenerate qubits ($H\equiv0$, $V\equiv1$); (b) entangled non-degenerate qubits, ($G\equiv0$, $L\equiv1$ in arm 1; $G\equiv0$, $R\equiv1$ in arm 2); and (c) entangled non-degenerate qutrits, ($L\equiv0$, $G\equiv1$, $R\equiv2$), where every second row is labelled. For all three cases, imaginary components were $<$0.03.}
\label{tomoResults}
\end{figure}

The simplest degenerate qubit encoding has first order logical basis states, e.g., $HG_{10}\equiv0$, $HG_{01}\equiv1$. The corresponding physical measurements required for tomography are then the states described by Padgett \emph{et al.} \cite{PadgetMJ1}: the $HG_{01}$-type modes with horizontal ($H$), vertical ($V$), diagonal ($D$) and anti-diagonal ($A$) phase discontinuities, and the $LGV_{0,\pm 1}$ modes with charge $\pm$1 phase singularities (right, $R$, and left, $L$). These states are measured using 6 different plane-wave hologram segments as shown in Fig.~\ref{dcQST}(d) \cite{note}. To test the performance of the SMA, we holographically created and measured a range of single qubit states using a coherent source (10 mW HeNe laser). In all cases, we obtained extremely high purities ($>$0.999) and fidelities with their ideal counterpart ($>$0.98). Fig.~\ref{tomoResults}(a) shows the two-photon state of the down-converter measured in the qubit basis: the state is highly entangled, the fidelity with the maximally-entangled $\phi^{+}$ Bell state is $F_{\phi^{+}}$=0.97. The degree of entanglement and mixture of the measured state is quantified, respectively, by the \emph{tangle}, $T$=0.90, and \emph{linear entropy}, $S_L$=0.06 \cite{WhiteAG2}.

The simplest non-degenerate qubit encoding has zero ($G\equiv0$) and first order (e.g. $R$ or $L$ $\equiv1$) basis states. The basis states are measured with the appropriate hologram segments, and the superposition states are simply accessed by displacing the $R$ or $L$ singularity a distance $\omega/\sqrt{2}$ from the centre of the beam [Fig.~\ref{dcQST}(e)], where $\omega$ is the intensity $1/e^2$ point \cite{VaziriA2,note1}. The analyser quality is equivalent to the degenerate case. The measured non-degenerate, two-qubit state [Fig.~\ref{tomoResults}(b), $T$=0.65 and $S_L$=0.11] has a lower tangle, reflecting the larger component of $G$ in the down-conversion beam. This state has a high fidelity, $F$=0.95, with a nonmaximally entangled state \cite{WhiteAG1} of the form $(|GG\rangle+\varepsilon|LR\rangle)/\sqrt{1+\varepsilon^2}$ for $\varepsilon$=0.60. The results for both types of qubit indicate that a Bell inequality could be violated \cite{MunroWJ1}.

We now encode a non-degenerate qutrit using basis states from the lowest two mode orders \cite{VaziriA1}: $L\equiv0$, $G\equiv1$ and $R\equiv2$. Our two-state tomographic technique enabled us to use the hologram in Fig.~\ref{dcQST}(d) and \ref{dcQST}(e); the resulting measured two-qutrit state is shown in Fig.~\ref{tomoResults}(c). This state is quite pure, with linear entropy $S_L$=0.18, and highly entangled. There are several ways to quantify the entanglement of this state. Given the relative populations of the basis states, we expect a non-maximally entangled state of the form, $(|LR\rangle+\varepsilon|GG\rangle+|RL\rangle)/\sqrt{2+|\varepsilon|^2}$; for $\varepsilon$=$1.79 e^{-0.07 i \pi}$, found using numerical optimisation, the fidelity between the ideal and measured nonmaximally entangled states is $F$=0.88. More directly, we calculate an upper bound to the measured \emph{entanglement of formation} of 0.74 \cite{BennetCH1,note2}.

One advantage that entangled qutrits offer over qubits is increased security in cryptographic protocols such as quantum bit commitment (BC) and coin flipping. Quantum BC binds a sender (Alice) to one message (a bit), and prevents the receiver (Bob) from determining the message before Alice later chooses to reveal it. BC is the basis for the most secure known strong quantum coin-flipping protocols \cite{SpekkeRW1}. While BC protocols with unconditional security are impossible \cite{LoHK,MayersD}, they can be partially secure \cite{SpekkeRW1}. The best known BC protocols are purification protocols, where Alice supplies the only quantum system, which consists of two parts. She sends the \emph{token} subsystem to Bob to commit her bit and the \emph{proof} subsystem later to reveal it. Maximum security in such protocols can be achieved by using two entangled qutrits for the token and proof, but not qubits.

\begin{figure}[b!]
\includegraphics[width=80mm]{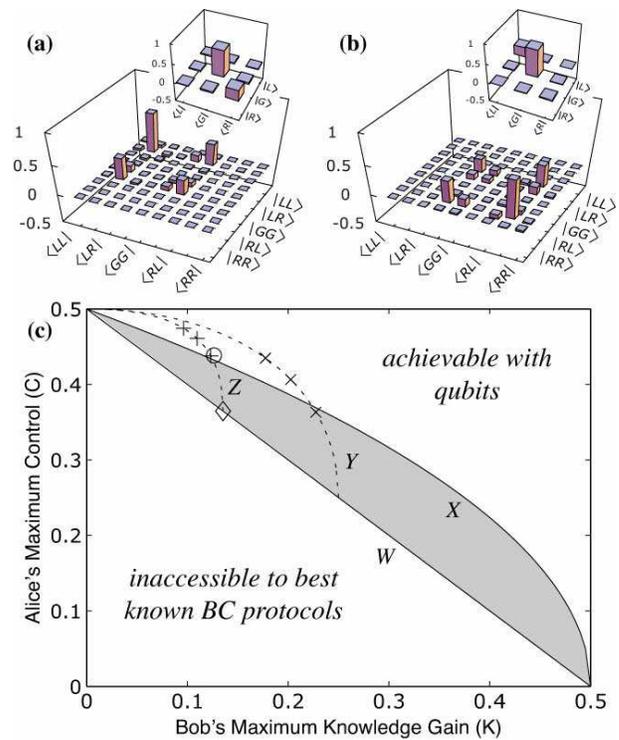}
\caption{ A purification bit-commitment (BC) protocol. The logical bits generated by Alice as described in the text: (a) $|0\rangle_L$; (b) $|1\rangle_L$. Insets: Bob's reduced density matrices -- the token subsystems. (c) Plot of Alice's Control vs Bob's Knowledge Gain. $\color[rgb]{1,0,0} \circ \color[rgb]{1,0,0}$: the measured protocol; $ \color[rgb]{0,0,1} \Diamond  \color[rgb]{0,0,1}$: the closest ideal protocol. \emph{W} and \emph{X}: the best known qutrit and qubit protocols. \emph{Y} \& \emph{Z}: Imperfect purification protocols with token states of the form, $\rho_{0,1} = p/3 \ I + (1-p) \ \rho_{0,1}^{ideal}$, where \emph{Y} is $\lambda$=0.5 and \emph{Z} is $\lambda$=0.27. The positions for $p=0.09,0.19,0.29$ are marked with $\times$ (\emph{Y}) and $+$ (\emph{Z}).}
\label{bcResults}
\end{figure}

We now outline one procedure for using our entangled qutrit state analysed above to implement a purification BC protocol. Depending on her choice of bit, Alice should prepare two qutrits in one of the 
orthogonal states $|0\rangle_L = \sqrt{\lambda} |12\rangle \!\!+\!\! e^{i\phi} \sqrt{1-\lambda} |01\rangle$ or $|1\rangle_L = e^{i\phi} \sqrt{1-\lambda} |21\rangle + \sqrt{\lambda} |10\rangle$, where $\lambda$ is a parameter characterising the security of the protocol. To prepare such states using our system, Alice needs to postselect the entangled states that have no photons in one of the basis modes of one subsystem: e.g., consider the proof subsystem in arm 1: zero photons in the ``2'' basis mode yield $|0\rangle_L$; zero photons in the ``0'' mode yield $|1\rangle_L$. In principle, manipulating the individual modes of the proof subsystem can be accomplished using a holographic interferometer in that arm. Postselection would then require either perfect detectors or spatial-mode quantum nondemolition (QND) measurements. Here, however, we simulate this process and reconstruct the new states \cite{qbcnote}. The logical states are then created by swapping the remaining proof subsystem modes. Fig.~\ref{bcResults}(a) and \ref{bcResults}(b) show the two-qutrit logical states that result from this simulated state preparation step. In this simulation, the only imperfections in the protocol arise from the initial state, thus giving a bound for the usefulness of our entangled qutrits. 

After preparing the appropriate state, Alice then sends the token qutrit to Bob. Because of the entanglement (quantified by $\lambda$), the reduced token state possessed by Bob is mixed, which lies at the heart of the security of the purification protocol. The fact that \emph{orthogonal} two-qutrit logical states produce \emph{non-orthogonal} token states provides some security against Bob cheating. His maximum knowledge gain, $K$, is limited by the distinguishability of these states and quantified by the trace distance. However, it is this partial distinguishability which in turn limits Alice's ability to cheat and change her bit after her commitment. Her maximum control, $C$, can be quantified by the fidelity between the token states. Details can be found in Ref.\ \cite{SpekkeRW1}. The protocol is concluded by Alice sending the proof qutrit to Bob, who performs the orthogonal, two-qutrit projective measurement, and either decodes the bit $\lbrace$$|0\rangle_L\langle0|$, $|1\rangle_L\langle1|$$\rbrace$ or catches Alice cheating. 

Figure~\ref{bcResults}(c) shows a plot of $C$ vs $K$, where the bottom left corner represents unconditional security and the top right corner represents no security. The ideal token states for this scheme give $K$=$\lambda/2$ and $C$=$(1-\lambda)/2$, and varying $\lambda$ produces the best known Alice-supplied security curve (\emph{W}). The shaded region between \emph{W} and \emph{X} highlights the area inaccessible to qubit-based, but accessible to qutrit-based BC protocols. The insets to Fig.~\ref{bcResults}(a) and (b) show the reduced density matrices for the token resulting from our initial state, which are closest to ideal states with $\lambda$=0.27 ($F$$\sim$0.99). However, in spite of this high fidelity, if we determine $C$ and $K$ directly from the measured token states, the protocol lies just inside the area accessible to qubits: a direct result of the slight ($<$3\%) residual population in the other mode of Bob's token subsystems, originating from the defects of Alice's original state. In other words, a two-qutrit state with residual populations of $<$1\% is required to surpass the qubit boundary (\emph{X}).

To implement this BC protocol, Alice must be able to perform deterministic postselection (e.g., using QND measurements). This is hard. Even if she achieves this perfectly, we have shown that the protocol still lies in the qubit-accessible regime. In our simulation, the only differences between our protocol and the ideal resulted from imperfections in the initial state. This result demonstrates that the requirements on the initial two-qutrit entangled state are extremely stringent, and that future theoretical work in this area should consider the critical role of even small amounts of mixture.

We have performed the first full characterisation of entangled, spatially-encoded quantum states, and achieved the first complete measurement of an entangled, two-qutrit state in {\em any} encoding, using a novel quantum tomography technique that only requires two-state superpositions. We have outlined a scheme for using this system to implement the best known BC protocol. With this measured state, this protocol would not reach maximal security, but we can see from the results what improvements are required. This analysis would have been impossible without access to the complete two-qutrit state, gained through quantum tomography.

This work was supported in part by the ARC, and by the MURI Center for Photonic Quantum Information Systems, ARO/ARDA program DAAD19-03-1-0199. A.~G. acknowledges support from the NZ FRST, J.~L.~O'B. and G. J. P. from the ARC COE CQCT.


\begin{thebibliography}{16}
\bibitem{note0} Violating Bell's inequality is not a measure of entanglement, as states with different amounts of entanglement and mixture can give the same Bell violation \cite{MunroWJ1}.
\bibitem{note} Plane-wave holograms do not produce/analyse single Gaussian modes, e.g. $|LGV_{0,-1} \rangle$. Instead they yield Gaussian-mode superpositions, e.g. $\sum^{\infty}_{i=0} c_{i} |LGV_{i,-1} \rangle$, where $c_{i}$$>$$c_{i+1}$ --- a weighted sum of modes of the same charge (--1), but with increasingly greater spatial extent and number of circumferential rings. A good approximation to single mode behaviour can be obtained by appropriate normalisation of the Stokes' parameters used in tomography, i.e. $\mathcal{S}_j = (\mathcal{N}_X - \mathcal{N}_Y)/(\mathcal{N}_X + \mathcal{N}_Y)$, where $(j,X,Y)$ is $(1,H,V)$, $(2,D,A)$, or $(3,R,L)$, and $\mathcal{N}_X$ is the number of counts in the $X$ basis. If exact behaviour is required, then Gaussian holograms should be used.
\bibitem{note1} This should not be done for higher-order, $|l|$$>$1, modes. Displacing a charge $|l|$$>$1 hologram to measure non-degenerate qudit superpositions -- as was done in \cite{MairA1,VaziriA3} -- analyses a superposition of several orders, not just the desired superposition of $|\zeta \rangle =(|LGV_{00} \rangle + |LGV_{0,\pm l} \rangle)/\sqrt{2}$. This is apparent in the output intensity pattern: a displaced charge $|l|$ hologram produces a mode with a single displaced vortex \cite{MairA1}; whereas a hologram for $|\zeta \rangle$ will produce/analyse a mode with $|l|$ single-charge vortices \cite{SoskinMS1}.
\end{thebibliography}
\end{document}